# Langmuir mechanism of low-frequency stimulated Raman scattering on nanoscale objects

V.B. Oshurko[1,2], O.V. Karpova[3], A.N.Fedorov[1], M.A. Davydov[1], A.F. Bunkin[1], S.M. Pershin[1], M.Ya. Grishin[1]

[1] *Prokhorov General Physics Institute of the Russian Academy of Sciences*
[2] *Moscow State University of Technology "STANKIN"*
[3] *Lomonosov Moscow State University*
*vbo08@yandex.ru*

**Abstract.** A simple physical mechanism of stimulated light scattering on nanoscale objects in water suspension similar to Langmuir waves mechanism in plasma is proposed. The proposed mechanism is based on a dipole interaction between the light wave and the non-compensated electrical charge that inevitably exists on a nanoscale object (a virus or a nanoparticle) in water environment. The experimental data for tobacco mosaic virus and polystyrene nanospheres are presented to support the suggested physical mechanism. It has been demonstrated that stimulated amplification spectral line frequencies observed experimentally are well explained by the suggested mechanism. In particular, the absence of lower frequency lines and the generation lines shift when changing the pH are due to ion friction appearing in the ionic solution environment. The selection rules observed experimentally also confirm the dipole interaction type. It has been shown that microwave radiation on nanoscale object acoustic vibrations frequency should appear under such scattering conditions. We demonstrate that such conditions also allow for local selective heating of nanoscale objects by dozens to hundreds degrees K. This effect is controlled by the optical irradiation parameters and can be used for affecting selectively certain types of viruses.

**Introduction**

Recently, many works have been published on direct observation of acoustic vibrational modes of nanoscale objects (in particular, viruses) by means of low-frequency optical Raman scattering [01] and extraordinary acoustic Raman scattering (EAR) [02,03]. However, these techniques allow to study only single particles placed in artificial environment, e.g. on a surface or using double nanoholes [03], thus limiting practical applications. A conventional electrostriction mechanism is suggested for all these cases, but there are many publications concerning low-frequency stimulated Raman scattering on viruses and nanoparticles in water environment [1-3] that are based on a theory of a stronger excitation mechanism analogous to Langmuir waves in plasma. As it turns out, this mechanism explains several ambiguous features of such light scattering. Obviously, of greatest interest are nanoscale biological objects like viruses in water suspensions [4]. There is no doubt that gaining information about such objects by a new technique that also provides an opportunity to affect these objects is important for biomedical applications. The suggested new mechanism opens a possibility of selectively affecting certain types of viruses by optical means, namely biharmonic pumping.

At the same time, the mechanism of stimulated light scattering is not completely clear for such objects. Stimulated light scattering with frequency shift corresponding to acoustic vibrations frequency characteristic for nanoscale objects is traditionally explained by models based on electrostriction (ponderomotive) mechanism. This mechanism suggests the modulation of electric susceptibility of an electrically neutral medium by an external field [3] which clearly does not fit the discussed case. It is known that biological and non-biological objects in water suspensions (e.g. viruses and polymeric nanospheres, respectively) have a significant non-compensated electrical charge [5] and virus' protein shell (capsid) composed of amino acids undergoes electrolytic dissociation in water environment [6]. This explains the phenomenon of proteins and ribonucleic acids electrophoresis caused by non-compensated electrical charge as the number of amino acid residues or nucleic bases is different from the number of alkaline residues [6]. Consequently, proteins and viruses demonstrate a so called isoelectric point, i.e. such pH of the environment when the non-compensated charge vanishes. It means that by varying the pH of the solution one can control the virus non-compensated charge value.

The published experimental works show that polymeric nanospheres in water suspensions also possess a significant charge, e.g. 20e (elementary charges) per single 50-nm polystyrene nanoparticle [5]. It is obvious that the energy of electrostatic interaction between such charge and the electric field in the laser beam waist is many orders of magnitude larger that the electrostriction interaction energy (i.e. the energy of interaction between the induced dipole moment and the external field). The electrostatic force affecting the charge is evidently much stronger than electrostriction forces. Until now, this scenario has not been taken into account, although the interaction between the light beam electromagnetic field and the charge may be the main mechanism determining low-frequency stimulated light scattering phenomenon.

Moreover, the existence of a significant non-compensated electrical charge leads to the strong suppression of acoustic vibrations in such particles surrounded by water due to so called ionic (or cataphoretic) friction [5]. In other words, the friction should be so strong that all vibrations were virtually impossible, however, the experiments show that exactly in water suspensions a directional stimulated emission is formed on Stokes frequency (i.e. red-shifted by the value of acoustic vibrations frequency inherent to these particles). Here and on we will call this effect the "low-frequency stimulated light scattering" (LFSS).

It should be noted additionally that there is a series of LFSS peculiarities compared to conventional stimulated Raman scattering (SRS). For example, unlike SRS, LFSS is almost never observed on the frequency of the lowest vibrational mode [1-3]. Also, as it will be shown later, for the case of viruses the observed LFSS frequency can depend on ions and virus concentration in the solution.

On the other hand, LFSS differs noticeably from the stimulated Brillouin scattering (SBS) process as well. In SBS, the stimulated light scattering appears on an acoustic wave inside a macroscopic object and the frequency shift is defined only by the speed of sound in this object. In turn, in LFSS the frequency shift is clearly caused by the scattering on acoustic waves in nanoscale objects and depends on these objects' size and properties.

Summing up, the above-mentioned peculiarities and unclear mechanism of the stimulated light scattering on nanoscale objects' acoustic vibrations allow to state that LFSS is a new nonlinear optics phenomenon.

The present work is aimed at developing a model describing the LFSS phenomenon based on a charge mechanism and at explaining the observed features of the process.

**Charge mechanism of LFSS**

Let us consider a simplified model of a charged dielectrical nanoscale object (nanosphere or a nano-sized virus particle) in the electromagnetic field of a light wave. Important is that the object itself is not electrically neutral in this case. Usually, the excess charge of such objects is compensated by oppositely charged ions of the solvent or the surfactant which is always added to the nanoparticle suspension to prevent adhesion. It is clear that this charge-compensating ionic shell size is defined by the Debye length for the solution. The existence of such shell requires to add another factor to the model, namely the friction that inevitably appears during acoustic vibrations of a nanoscale object in an ionic shell.

Let the plane-polarized light wave be directed along the z axis and the electric field vector be oscillating along the x axis. The Maxwell equations for the light wave $\vec{E}(\vec{r},t)$ give

$$\hat{D}\vec{E}(\vec{r},t) = \nabla(\nabla\vec{E}(\vec{r},t)) + \frac{4\pi}{c^2}\frac{\partial \vec{J}(\vec{r},t)}{\partial t} \qquad (1)$$

where $\hat{D} = \Delta - 1/c^2 \partial^2/\partial t^2$ is d'Alembertian, $c$ is the speed of light and $\vec{J}(\vec{r},t)$ is the density of current induced by the acoustic movement of charge. Note that in the case of regular SRS or ponderomotive mechanism the first term in the right part equals zero due to zero total charge of the medium. It is not so in our case and the field divergence equals the charge density:

$\nabla \vec{E}(\vec{r},t) = 4\pi\rho(\vec{r},t)$, where $\rho(\vec{r},t)$ is the charge volume density.

Let the object be charged uniformly with the initial charge density constant over the volume: $q_0 = Const$. If the object is absolutely rigid (its elastic modulus equals infinity), it is obvious that there will be no movement apart from reciprocating oscillations in the field $\vec{E}$ of a light wave. Such oscillations make no contribution to inelastic light scattering and will not be further taken into account. Otherwise, acoustic vibrations are possible in an elastic object with a finite Young's modulus and can appear spontaneously at room temperature because acoustic phonon energy $\hbar\Omega$ is much less than the heat energy $k_B T$.

Let us consider the acoustic displacement $\vec{u} = \vec{r}' - \vec{r}$ where $\vec{r}'$ is the displaced position of the point $\vec{r}$ in an acoustic wave. Solid body acoustic equation can be expressed as

$$\Delta \vec{u}(\vec{r},t) - \Gamma \frac{\partial \vec{u}(\vec{r},t)}{\partial t} - \frac{1}{v^2}\frac{\partial^2 \vec{u}(\vec{r},t)}{\partial t^2} = \frac{1}{v^2}\vec{F}_m \qquad (2)$$

where $v$ is the speed of sound and $\vec{F}_m$ is the force volume density ($\vec{F}_v = d\vec{F}/dV$) divided by the density of mass, i.e. $\vec{F}_m = \vec{F}_v/\rho$. Here, we introduce phenomenologically the term describing the friction proportional to velocity with the coefficient $\Gamma$. As it was mentioned earlier, this is necessary to account for external ionic shell resistance to oscillations.

It is obvious that in our case the force density is $\vec{F}_m = q(\vec{r},t)/\rho(\vec{r},t)\vec{E}(\vec{r},t)$, where $q$ is the charge volume density. An acoustic wave in nanoscale object can be naturally considered as a wave of density modulation $\rho'(\vec{r},t) = \rho_0 f(\vec{r},t)$, where $f$ is the function describing this wave. The charge density modulation is described by the same function $f$ and, consequently, the ratio $q'/\rho'$ in the right part of Eq. (2) remains constant.

Then, the only way of direct interaction between the light wave field and nanoscale object charge is "charge decompensation" caused by a well-known Debye-Falkenhagen effect [7]. When an electromagnetic wave passes through ionic solution, starting from some wave frequency the ion "heavy" hydration shell fails to "keep up" with the ion movement. It leads to a sharp growth in the solution conductivity. For the majority of water solutions this "starting" frequency is about several GHz which is lower than the observed characteristic frequencies of nanospheres/viruses acoustic vibrations (typically dozens of gigahertz). In recent research, such inertia of hydration shells is always taken into account [8] to explain the appearance of non-compensated charge while exciting acoustic vibrations in charged nanoscale objects.

It is easy to see that the nature of "charge decompensation" wave is analogous to the well-known Langmuir waves in plasma [9] with the only difference that the "charge decompensation" wave characteristics depend on nanoscale object's mechanical properties.

For simplicity, consider a one-dimensional problem describing a tobacco mosaic virus particle which looks like a thin cylinder. Let the cylinder axis be directed along the $x$ axis like the light wave $E$ vector. If the virus' negative charge volume density is modulated by an acoustic wave $q^-(x,t) = q_0^- f(x,t)$, then the opposite ion charge density in the solution remains constant $q^+ = q_0^+$ (due to the Falkenhagen effect) and equals $q_0$ (for the one-dimensional problem). Then the total charge density is defined by the difference $q(x,t) = q_0^- f(x,t) - q_0^+$. An easy direct calculation shows that the total charge density is $q(x,t) = q_0 \frac{\partial u}{\partial x}$.

Apart from the force induced by the light wave field $q_0 u_x E(z,t)$, there are also forces of interaction between charged areas similarly to that for Langmuir waves in plasma. These forces are

$$F_e = q(x,t)E' = q(x,t)\int \frac{\partial E'}{\partial x}dx = q(x,t)\int 4\pi q(x,t)dx = 4\pi q_0^2 u u_x \qquad (3)$$

where $E'$ is the field created by non-compensated charges.

Finally, the equation for acoustic displacement $u(x,t)$ is:

$$u_{xx} - \Gamma u_t - \frac{1}{v^2}u_{tt} = \frac{q_0}{\rho_0 v^2}u_x E(z,t) + 4\pi \frac{q_0^2}{\rho_0 v^2}u u_x \qquad (4)$$

**Dipole approximation**

It is rather complicated to find a general solution of this equation for a harmonic light wave, however, it is clear beforehand that a charge moving in the field of a light wave does not itself cause inelastic scattering. To find inelastic components of the scattering, it is necessary to examine the processes in which the light wave field modulates some parameter of the system. In particular, let us consider the modulation of the total dipole moment by the external electric field. To achieve this, an auxiliary problem has to be solved: consider a constant force affecting the object's dipole moment instead of a light wave field, i.e. let us determine how a constant external field changes the dipole moment.

In the right part of Eq. (4) there is density of the force affecting the object via the field, and the interaction between the areas of the distributed charge. The estimation readily shows that although the LFSS field is relatively weak compared to the left part of Eq. (4), the second term in the right part of Eq. (4) (namely, the interaction between the areas) is much smaller than the first term for realistic acoustic vibration amplitudes and can be neglected.

It is obvious that in the absence of acoustic vibrations the charge distribution is uniform, and there is no dipole moment (or, to be precise, it is possible to choose a reference frame in wich the dipole moment equals zero). The dependence of a charged object dipole moment on the reference frame choice is insignificant as force is defined only by the dipole moment derivative. When an acoustic wave appears, charged areas are created due to charge decompensation, and the total dipole moment (per unit volume) for a virus of the length $L$, and its derivative, are

$$p = \frac{q_0}{L}\int_0^L x u_x dx \qquad (5)$$

and

$$p_x = \frac{q_0}{L}L u_x(L) = q_0 u_x(L)$$

In dipole approximation, to determine the force affecting each element $dx$ in the point $x$ one may determine the coordinate-dependent dipole moment $p(x)$ and its derivative as

$$p_x(x) = q_0 u_x$$

The according light wave electric field force density is

$$F_v = \frac{1}{2}p_x E(z,t) = \frac{1}{2}q_0 u_x E(z,t) \qquad (6)$$

Estimations show that the dipole moment magnitude and the force affecting the dipole are in fact much bigger than electrostriction (ponderomotive) forces traditionally considered in nonlinear optics.

In (5) and (6) the displacement $u(x,t)$ and, consequently, the dipole moment are a function of the external field E. Consider the dependence $p(E)$ using the simplified model. Let us examine a charged object (a virus) in a constant external field $E$. Now, $E = Const$ in the right part of Eq. (4). Then, using a standard substitution $u(x,t) = U(t)V(x)$ we get a system of equations

$$U_{tt}(t) + \gamma U_t(t) + \Omega_m^2 U(t) = 0 \qquad (7)$$

$$V_{xx}(x) - \frac{q_0}{2\rho_0 v^2} E V_x(x) + \frac{\Omega_m^2}{v^2} V(x) = 0 \tag{8}$$

(here, we designate $\gamma = \Gamma v^2$ and the common constant $-\Omega_m^2/v^2$). The solution of Eq. (8) with the free ends condition $u_x(0) = u_x(L) = 0$ is

$$V(x) = u_0 e^{\frac{q_0 E x}{4\rho_0 v^2}} \cos\left(x \sqrt{\frac{\Omega_m^2}{v^2} - \frac{E^2 q_0^2}{16\rho_0^2 v^4}}\right) \tag{9}$$

where $\Omega_m = m\pi v/L$ is the $m$-th mode natural frequency and $u_0$ is the amplitude. From here, one can determine the derivative of dipole moment according to (5), (6) and (9):

$$p_x(E) = U(t) \frac{u_0 q_0}{4\rho_0 v^2} e^{\frac{q_0 E L}{4\rho_0 v^2}} \left(q_0 E \cos(\Omega_E L) - (4v^2 \rho_0 \Omega_E)\sin(\Omega_E L)\right) \tag{10}$$

where

$$\Omega_E = \sqrt{\frac{\Omega_m^2}{v^2} - \frac{E^2 q_0^2}{16\rho_0^2 v^4}} \tag{11}$$

Here, $U(t)$ is the solution of (7) in the form of a periodic function with $\sqrt{\Omega_m^2 - \gamma^2}$ frequency and attenuation coefficient $\gamma$. For our purposes, let $U = 1$. We emphasize that in Eqs. (7) and (8) the field $E$ is constant and the Eqs. (7), (8) and (9) are not meant to solve the stimulated amplification problem but are only used to estimate the dipole moment dependence on the field.

In the conditions of LFSS the field can be considered small $E << 4\rho_0 \Omega_m v/q_0$. This is true up to $E:10^9$ V/m which is significantly stronger than intensities used in experiments $I:1$ MW/cm². Then, $p_x(E)$ can be expanded into a series by $E$:

$$p_x(E) = \frac{q_0^2 u_0 (-1)^m}{4\rho_0 v^2} E + \frac{3}{32} \frac{q_0^3 u_0 L (-1)^m}{\rho_0^2 v^4} E^2 + \ldots \tag{12}$$

It is important to note that the series coefficients depend on the initial oscillations amplitude $u_0$ which is not so for the regular molecular SRS. In fact, almost any molecules have non-zero polarizability, however, in our case it appears only when non-uniform charge distrinution is already present due to, for example, spontaneous acoustic vibrations. If there is no such non-uniform distribution, then the polarizability, i.e. the dipole moment dependence on external field, evidently equals zero: the external field itself does not create the non-uniform charge distribution (and, consequently, the dipole moment) and affects all the areas equally.

**Stimulated amplification of radiation**

Now, given the dipole approximation, substitute the force density with $1/2 p_x(E(z,t))E(z,t)$ in the right part of the acoustic equation (4). Accoynting for the expansion (12), for the first two terms of the series we get

$$u_{xx} - \Gamma u_t - \frac{1}{v^2} u_{tt} = \alpha_1 E(z,t)^2 + \alpha_2 E(z,t)^3 \tag{13}$$

where

$$\alpha_1 = \frac{q_0^2 u_0 (-1)^m}{2\rho_0 v^2}, \alpha_2 = \frac{3 q_0^3 u_0 L (-1)^m}{16\rho_0^2 v^4} \tag{14}$$

and $E(z,t)$ is the light field. To determine the amplification of radiation components in a stimulated process, consider a combination of a pump wave and a Stokes wave according to traditional nonlinear optics approach $E(z,t) = E_0 exp(i\omega_0 t - ik_0 z) + E_s exp(i\omega_s t - ik_s z) + c.c.$. In this case the right part of (13) does not depend on $x$, and (13) can be divided to two ordinary equations

$$U_{tt}(t) + \gamma U_t(t) + \Omega_m^2 U(t) = \alpha_1 E(z,t)^2 + \alpha_2 E(z,t)^3 \qquad (15)$$

and

$$V_{xx}(x) + \frac{\Omega_m^2}{v^2} V(x) = 0 \qquad (16)$$

The second equation (16) with the free ends condition gives an expression for natural frequencies $\Omega_m = m\pi v/L, m = 1,2,..$

$$V(x) = u_0 \cos\left(\frac{\Omega_m}{v} x\right) \qquad (17)$$

In the second equation (15), substitute the above-mentioned combination of pump and Stokes waves (frequencies $\omega_0$ and $\omega_s$, respectively) and pick out a solution $u = UV$ at the frequency $\Omega = \omega_0 - \omega_s$:

$$u_\Omega = \alpha_1 V(x) \frac{E_0 E_s}{\Omega_m^2 - \Omega^2 + 2i\gamma\Omega} e^{(i\Omega t - ik_\Omega z)} \qquad (18)$$

Similarly, pick out a solution at the Stokes frequency $\omega_s$:

$$u_s = \alpha_2 V(x) \frac{E_0^2 E_s}{\Omega_m^2 - \omega_s^2 + 2i\gamma\omega_s} e^{(i\omega_s t - ik_s z)} \qquad (19)$$

It can be seen that the second solution for the Stokes component $\omega_s$ does not have pronounced resonances at detuning equal to natural frequencies $\omega_0 - \omega_s = \Omega_m$ and does not explain the observed stimulated amplification at these frequencies.

Consequently, the only possible stimulated amplification mechanism for Stokes/anti-Stokes frequencies is a simultaneous amplification at a microwave frequency $\Omega$ (when waves with frequencies $\omega_0$ and $\omega_s$ are combined) and further amplification of $\omega_s$ frequency when $\omega_0$ and $\Omega$ frequencies are combined.

**Coupled waves equations**

To investigate the possibility of such subsequent amplification, consider the propagation of four waves at the frequencies $\omega_0, \Omega, \omega_s$ and $\omega_a$ (where $\omega_a$ is anti-Stokes radiation frequency). Like before, pick out the solutions of Eq. (13) for these frequencies taking into account only first order terms in the right part of the equation:

$$u_0 = \alpha_1 V(x) \frac{(E_\Omega E_s + E_0 E_a)}{\omega_m^2 - \omega_0^2 + 2i\gamma\omega_0} e^{(i\omega_0 t - ik_0 z)} \qquad (20)$$

$$u_s = \alpha_1 V(x) \frac{E_0 E_\Omega}{\Omega_m^2 - \omega_s^2 + 2i\gamma\omega_s} e^{(i\omega_s t - ik_s z)} \qquad (21)$$

$$u_a = \alpha_1 V(x) \frac{E_0 E_\Omega}{\Omega_m^2 - \omega_a^2 + 2i\gamma\omega_a} e^{(i\omega_a t - ik_s z)} \qquad (22)$$

$$u_\Omega = \alpha_1 V(x) \frac{(E_0 E_s + E_0 E_a)}{\Omega_m^2 - \Omega^2 + 2i\gamma\Omega} e^{(i\Omega t - ik_\Omega z)} \qquad (23)$$

Now, for substitution to the wave equation (1), define the terms $J_t = \mu_0 \partial^2 u_i/\partial t^2$ and using slowly varying envelope approximation we get a system of bound waves equations (here we wave out the dependence on $x$ coordinate):

$$\frac{dE_0}{dz} = \beta_0 E_\Omega (E_s + E_a), \frac{dE_s}{dz} = \beta_s E_\Omega E_0$$

$$\frac{dE_a}{dz} = \beta_a E_\Omega E_0, \frac{dE_\Omega}{dz} = \beta_\Omega E_0 (E_s + E_a) \qquad (24\text{-}27)$$

where $\beta$ are the constants defined by (20)-(23). Solving these equations for the Stokes component we obtain

$$E_s(z) = \frac{E_{s0}}{1 - E_{s0} Re(\chi) z} \qquad (28)$$

where $E_{s0}$ is the magnitude of (e.g. spontaneous) Stokes radiation at $z = 0$ and $\chi$ is

$$\chi(\Omega) = \sqrt{-\frac{c^2 \mu_0^2 \alpha_1^2 \omega_0 \Omega}{(\Omega_m^2 - \omega_0^2 + 2i\gamma\omega_0)(\Omega_m^2 - \Omega^2 + 2i\gamma\Omega)}}$$

$$\left( \frac{(\omega_0 + \Omega)(\Omega_m^2 - (\omega_0 - \Omega)^2 + 2i\gamma(\omega_0 - \Omega))}{(\Omega_m^2 - (\omega_0 + \Omega)^2 + 2i\gamma(\omega_0 + \Omega))(\omega_0 - \Omega)} + 1 \right) \qquad (29, 30)$$

**Experimental results**

Now we are able to compare the developed theoretical model with the experimental results on low-frequency stimulated light scattering in suspensions of the tobacco mosaic virus (TMV) in tris buffer and polystyrene nanoparticles of different diameters (70 nm, 100 nm, 500 nm) and concentrations in distilled water.

In the experimental setup described in detail in [4] the cuvette with the studied suspension was irradiated by the focused second harmonic pulses of a single-frequency YAG:Nd$^{3+}$ laser (wavelength $\lambda = 532$ nm, pulse duration $\tau = 10$ ns, pulse energy $E_p$ up to 40 mJ). The radiation was focused into the cuvette center by a lens with focal length f=30 mm. Stimulated light scattering spectra were studied by a Fabry-Pérot interferometer and recorded by a complementary metal-oxide semiconductor (CMOS) camera Basler acA1920-40um.

The results of TMV suspension spectra measurements [11] are presented in Fig.5. For virus concentration: $1.0 \cdot 10^{12} cm^{-3}$ and laser pulse energy $E_p \sim 20$ mJ a LFSS line was detected at a 1.47 cm$^{-1}$ shift relative to the pump line, which corresponds to 44.1 GHz oscillation frequency (Fig.5, a), and for virus concentration: $2.0 \cdot 10^{12} cm^{-3}$ and laser pulse energy ~30 mJ a LFSS line was detected at 1.046 cm$^{-1}$ shift (31.38 GHz, Fig.5, b). The experiments also confirm the directional character of the LFSS emission.

Additionally, LFSS spectra were measured for suspensions of polystyrene nanoparticles of different diameters [12]. For particle diameter 74±11 nm and concentration in suspension $\sim 9 \cdot 10^{14} cm^{-3}$, spectral lines were observed at ~0.63 cm$^{-1}$ shift, which gives 18.9 GHz frequency. Particle diameter 540±11 nm and concentration $8 \cdot 10^7 cm^{-3}$ give the spectral line at ~0.67 cm$^{-1}$ shift (20 GHz frequency).

**Vibrational states selection rules**

Judging from the known (measured experimentally) values of Young's modulus and speed of sound in TMV [10], it is easy to calculate acoustic vibrations frequencies for a cylinder 300 nm in length and 18 nm in diameter in water environment. Numerical calculations performed in COMSOL Multiphysics software for all possible vibration types give the following results (supposing the speed

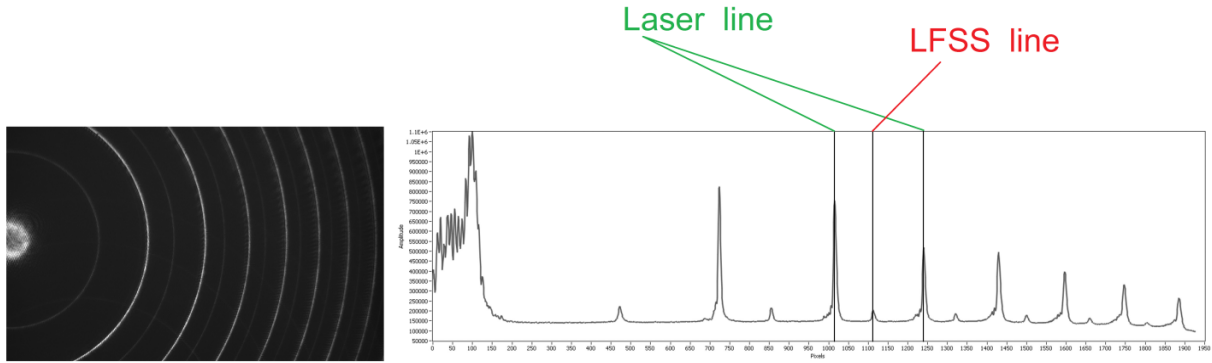

Fig.1, a) Low-frequency stimulated scattering (LFSS) spectrum for the tobacco mosaic virus concentration ~$1.0 \cdot 10^{12}$ cm$^{-3}$, $\Delta\nu$~1.47 cm$^{-1}$ (~44.1 GHz). Here and on "LFSS line" stands for the spectral line of low-frequency stimulated scattering (LFSS).

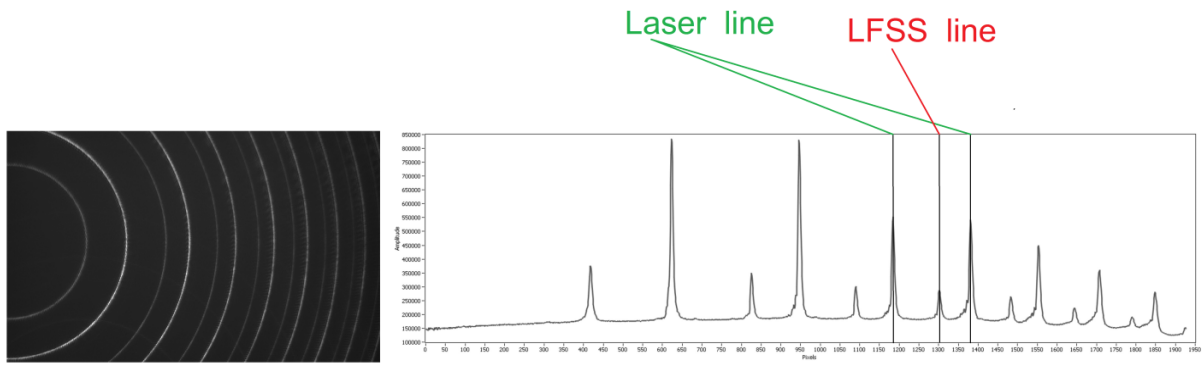

Fig.1, b) LFSS spectrum for the tobacco mosaic virus concentration ~$2.0 \cdot 10^{12}$ cm$^{-3}$, $\Delta\nu$~1.046 cm$^{-1}$ (~31.38 GHz).

of sound $v = 3430$ m/s, density $\rho_0 = 1100$ kg/m$^3$, Young's modulus E = 9.5 MPa [4]), see Table 1:

Table 1. TMV virus longitudinal vibration mode frequencies.

| m | 0 | 1 | 2 | 3 | 4 | 5 | 6 | 7 |
|---|---|---|---|---|---|---|---|---|
| f, GHz | 5.71 | 11.43 | 17.15 | 22.86 | 28.58 | 34.30 | 40.01 | 45.73 |

| m | 8 | 9 | 10 | 11 | 12 | 13 | 14 | 15 |
|---|---|---|---|---|---|---|---|---|
| f, GHz | 51.45 | 57.16 | 62.88 | 68.60 | 74.32 | 80.04 | 85.76 | 91.48 |

Bending vibration frequencies turned out to be higher than 112 GHz and radial vibration frequencies are above 190 GHz.

As it can be seen, longitudinal vibration frequencies 34 GHz and 45 GHz coincide well with the observed TMV LFSS frequencies (31 GHz and 44 GHz). The important thing to mention is that increasing the virus concentration in tris buffer in the experiments resulted in the observed LFSS frequency switch from 44 GHz to 31 GHz. In one experimental session, the lines 44 GHz and 31 GHz even appeared simultaneously, but the intermediate frequency $\approx 40$ GHz was never observed.

This fact is in good agreement with the suggested theoretical model. In fact, the $\approx 40$ GHz frequency corresponds to even vibrational mode for which the integral (5) defining the dipole moment according to (17) equals zero. Hence, the experiment confirms the dipole type of interaction and the subsequent selection rule for odd vibrational modes.

Now the absence of lower vibrational frequencies in the amplified LFSS lines can be explained. In fact, the expression (29) that denotes the gain increment defines only some "window" of amplification. As it can be seen from (29), this window depends on the magnitude of ionic friction $\gamma$. Fig.02 shows the gain increment $Re(\chi)$ as a function of $\Omega$ for a set of resonance odd mode frequencies $\Omega_m$ from Table 1 ($m = 1,3,..$) for two values of ionic friction coefficient (2.0 and 10.0 GHz). One can see that weaker friction (2.0 GHz) results in maximal gain increment $Re(\chi)$ for $m = 4$ at a frequency about 33 GHz. The evident reason for this is non-zero ionic friction. The increase of ionic friction coefficient to 10.0 GHz, as expected, leads to enormous broadening and merging of resonance peaks but the maximum appears at ~41 GHz.

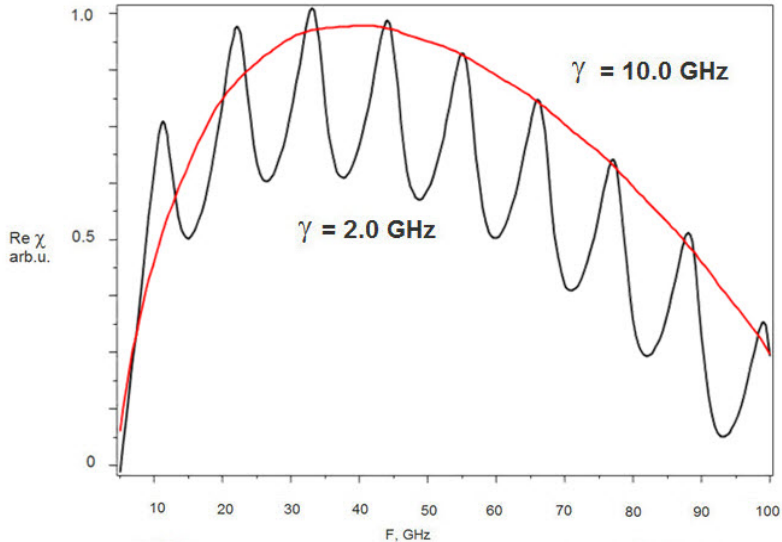

Fig.2. Gain "increment" for the frequencies of 10 odd modes (m=1,3,5,...) from Table 1 for two friction coefficients: 2.0 GHz (black) and 10.0 GHz (red).

Fig.03 and Fig.04 illustrate a growth of these components with the increase of interaction length $z$ according to (31). It can be seen that upon reaching $z = 2.6$ mm the spectrum is reduced virtually to a single line: 41 GHz line for the 10.0 GHz friction coefficient and 33 GHz line for the 2.0 GHz friction coefficient. Fig.02 shows the radiation spectra for $z = 2.6$ mm.

These results correspond well to the experimental data and allow for explaining the observed LFSS frequency change caused by the TMV concentration increase in the suspension. The ionic friction evidently decreases with the decrease in number of ions "compensating" the virus' own charge. During the experiments, we increased the virus concentration by evaporating water from the solution but "heavy" virus particles and tris molecules providing weak-alkaline environment were not evaporated, hence pH of the solution grew. As it was mentioned earlier, the isoelectric point (i.e. the pH for which the virus becomes electrically neutral) lies at rather high pH values (namely, alkaline). It means that with water evaporation the pH grows and non-compensated charge value decreases. Consequently, there are less solution ions compensating the charge of the virus leading to the reduction of ionic friction, which in turn results in amplification frequency change, as it can be seen from Fig.02-Fig.04.

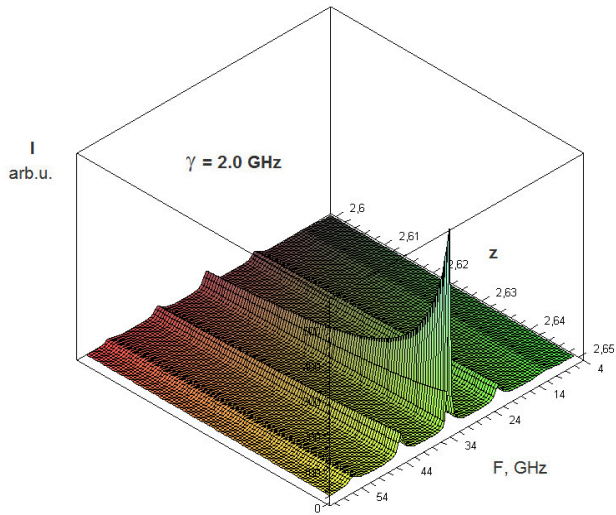

Fig.3. Selection of mode (lower friction): Gain dependence on the interaction length z for the frequencies of 10 odd modes (m=1,3,5,...) from Table 1 for the friction coefficient 2.0 GHz.

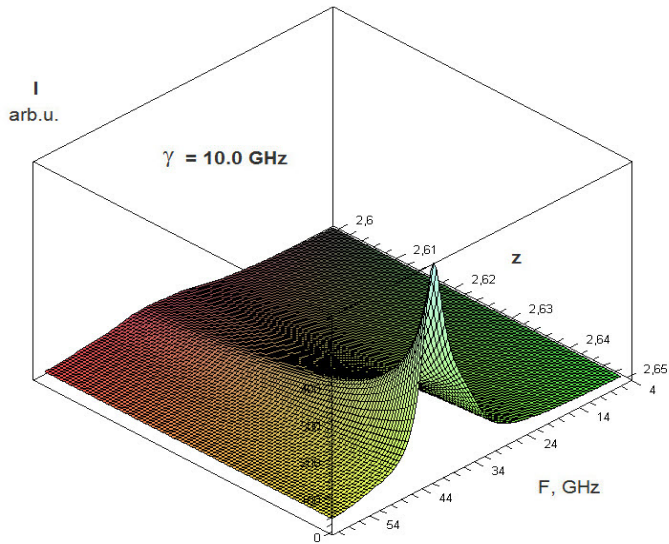

Fig.4. Selection of mode (higher friction): Gain dependence on the interaction length z for the frequencies of 10 odd modes (m=1,3,5,...) from Table 1 for the friction coefficient 10.0 GHz.

Note that only one adjustable parameter was used to describe precisely the LFSS phenomenon, namely the ionic friction magnitude. It is worth mentioning that the phenomenon can hardly be explained at all in terms of other physical mechanisms.

Another experimental result can be easily understood now. The experimental series revealed that for both 70-nm and 500-nm polystyrene particles suspensions LFSS lines with very close frequencies are observed, namely 18-20 GHz, although the particle size and consequently the lower resonance frequency differ by an order of magnitude. However, taking into account that friction does not depend on the surface area, the "window" of amplification is the same for the same friction coefficient for both cases according to (29), and only those modes are amplified whose frequencies lie around the amplification maximum.

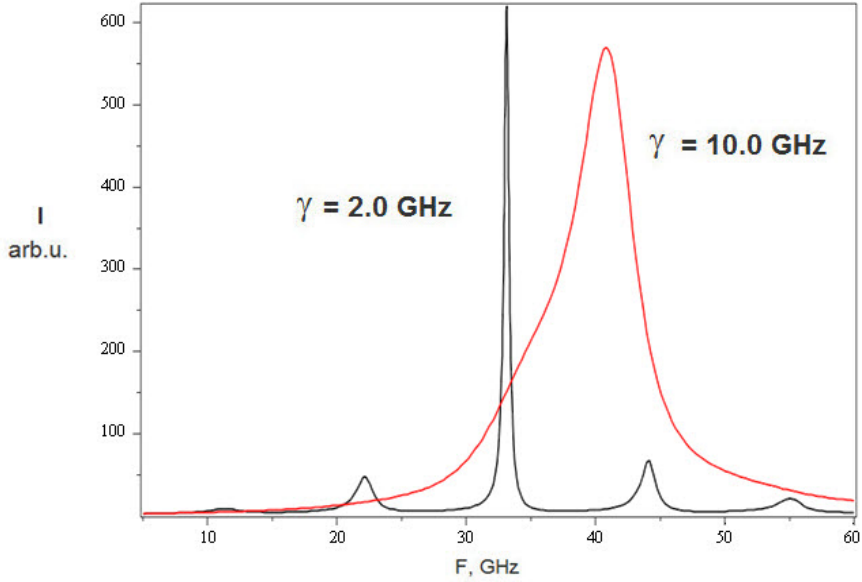

Fig.5. Result of selection: spectral lines at the exit from the interaction region z=2.5 mm with the frequencies of 10 odd modes (m=1,3,5,...) from Table 1 taken into account, for two friction coefficients: 2.0 GHz (black) and 10.0 GHz (red).

**Nanoscale objects heating**

Another conclusion follows directly from the prevoius. If a significant part of the pump energy is spent on overcoming the friction, this should result in local nanoscale particles heating. Let us estimate such heating roughly in terms of the described theoretical model. Suppose that the virus is a dipole with a moment $p \approx QL$, where $Q$ is non-compensated charge value experimentally estimated as $\sim 50 \cdot 1.6 \cdot 10^{-19}$ C. Then the energy of such dipole in the field $E$ is $A \approx QLE$ and the heat power can be taken as $P \approx QLE\Omega_m$. The part of this power turning into heat can be estimated from the ionic friction magnitude as $1 - exp(-\gamma\tau_p)$ where $\tau_p$ is the laser pulse duration. With the friction coefficient $\gamma \approx 10^9 s^{-1}$ and pulse duration $\tau_p \gg 1/\gamma$ all the absorbed radiation is transformed into heat. The amount of heat produced during the pulse $\tau_p$ is $A \approx QLE\Omega_m\tau_p$. Then, if we wave out thermal conductivity (during the laser pulse, heat wave diffusion length is ~100 nm given water thermal conductivity $\chi = 1.9 \cdot 10^{-5} cm^2/s$), we get the final heating temperature

$$T \approx \frac{QL\Omega_m\tau_p}{\rho_0 C_p V} E_0 \qquad (31)$$

where $C_p \approx 3000 J/(kgK)$ is protein heat capacity. Then, at the light field strength $E_0 \approx 10^8 V/m$ we get a heating by $T \approx 26.0 K$ during a 10 ns laser pulse. Hence, this rough estimate shows that in experiments heating can take place, however the temperature is still not sufficient for proteins denaturation.

Still, it is obvious that simple several-fold pulse duration increase (up to 0.1-1 μs) may lead to strong local heating up to dozens or hundreds degrees K. At the same time, overall heating may be insignificant. Note that the virus is heated by local selective non-resonant radiation that is not absorbed by the suspension. This is the stimulated nature of the process that provides highly selective heating of certain virus types.

It is known that tobacco mosaic virus heating only up to 94°C leads to partial capsid proteins denaturation and spherical virus formation [13,14]. This means that selective heating of given virus types that changes the capsid shape may strongly affect the virus functioning.

It is important that LFSS allows for selectively heating certain types of nanoscale objects up to substantial temperatures. The very possibility of such heating opens new prospects for local selective treating of nanoscale objects, viruses or cellular organelles by means of nonlinear optics for biomedical applications.

**Conclusion**

In the present work, a simple physical mechanism of low-frequency stimulated light scattering on nanoscale objects in water suspension was suggested. The suggested model differs significantly from conventional stimulated scattering mechanism. The model is based on a dipole interaction between the light wave and the non-compensated electrical charge that inevitably exists on a nanoscale object (a virus or a nanoparticle) in water environment. The experiments were carried out on observing low-frequency stimulated light scattering in tobacco mosaic virus and polystyrene nanosphere suspensions and the obtained data supports the suggested model. The selection rules observed experimentally also confirm the dipole type of the interaction. It has been demonstrated that stimulated amplification spectral line frequencies observed experimentally are well explained by the suggested mechanism. In particular, the absence of lower-frequency lines and the "window of amplification" are due to ion friction appearing in the ionic solution environment. It has been shown that under low-frequency stimulated light scattering conditions microwave radiation should appear at nanoscale object acoustic vibration frequency. We demonstrate that such conditions also allow for local selective heating of nanoscale objects by dozens to hundreds degrees K. This effect is controlled by the optical irradiation parameters and can be used for affecting selectively certain types of viruses.